\documentclass[10pt,twocolumn,letterpaper]{article}
\usepackage[accsupp]{axessibility}  

\usepackage{iccv}
\usepackage{times}
\usepackage{epsfig}
\usepackage{graphicx}
\usepackage{amsmath}
\usepackage{amssymb}
\usepackage{multirow}
\usepackage{multicol}
\usepackage{bbm}
\usepackage{float}
\usepackage{booktabs}
\usepackage{subfig}
\usepackage{algorithm} 
\usepackage{algorithmicx}
\usepackage{algpseudocode}

\usepackage[pagebackref=true,breaklinks=true,letterpaper=true,colorlinks,bookmarks=false]{hyperref}
\usepackage[capitalize]{cleveref}
\crefname{section}{Sec.}{Secs.}
\Crefname{section}{Section}{Sections}
\Crefname{table}{Table}{Tables}
\crefname{table}{Tab.}{Tabs.}
\iccvfinalcopy 

\def\iccvPaperID{5207} 

\ificcvfinal\fi
\newtheorem{theorem1}{Theorem}
\newtheorem{lemma1}{Lemma}
\newtheorem{proposition1}{Proposition}
\begin{document}

\title{Multi-metrics adaptively identifies backdoors in Federated learning}

\author{Siquan Huang$^1$ \\{\tt\small cssiquanhuang@mail.scut.edu.cn}
\and Yijiang Li$^{2}$ \\{\tt\small yli556@jhu.edu}
\and Chong Chen$^1$ \\{\tt\small cschenchong@mail.scut.edu.cn}
\and Leyu Shi$^1$ \\{\tt\small csshileyu@mail.scut.edu.cn}
\and Ying Gao$^1$\thanks{Corresponding Author} \\{\tt\small gaoying@scut.edu.cn}
\and $^1$School of Computer Science and Engineering, South China University of Technology\\
$^2$Johns Hopkins University
}


\maketitle
\ificcvfinal\thispagestyle{empty}\fi

\begin{abstract}
The decentralized and privacy-preserving nature of federated learning (FL) makes it vulnerable to 
backdoor attacks aiming to manipulate the behavior of the resulting model on specific adversary-chosen inputs. 
However, most existing defenses based on statistical differences take effect only against specific attacks, especially when the malicious gradients are similar to benign ones or the data are highly non-independent and identically distributed (non-IID).
In this paper, we revisit the distance-based defense methods and discover that i) Euclidean distance becomes meaningless in high dimensions and ii) malicious gradients with diverse characteristics cannot be identified by a single metric. To this end, we present a simple yet effective defense strategy with multi-metrics and dynamic weighting to identify backdoors adaptively. Furthermore, our novel defense has no reliance on predefined assumptions over attack settings or data distributions and little impact on benign performance. To evaluate the effectiveness of our approach, we conduct comprehensive experiments on different datasets under various attack settings, where our method achieves the best defensive performance. For instance, we achieve the lowest backdoor accuracy of $3.06\%$ under the most difficult Edge-case PGD, showing significant superiority over previous defenses. The experiments also demonstrate that our method can be well-adapted to a wide range of non-IID degrees without sacrificing the benign performance. 
\end{abstract}

\section{Introduction}
\label{sec:intro}

Federated learning (FL) \cite{konevcny2016federated,mcmahan2017communication} is a distributed machine learning paradigm that enables multiple participants to train a quality model collaboratively without exchanging their local data. 
During each round of training, the central server distributes the global model to a subset of clients. Each client updates the model with the local data and submits the parameters to the server for aggregation.
By training an efficient and quality model in a decentralized manner without sacrificing the privacy of participants, FL alleviates the possible conflict between technological developments and regulations for data privacy protection(\eg, General Data Protection Regulation). FL has already been applied and achieved success in multiple fields, such as image processing \cite{li2023diverse}, word prediction \cite{keyboard}, medical imaging \cite{li2022more}, and edge computing \cite{zhuang2021collaborative, xie2023weakly}.

However, FL is vulnerable to backdoors manipulating the model towards the targeted behavior on specific adversary-chosen inputs \cite{kairouz2021advances, bagdasaryan2020backdoor,bhagoji2019analyzing,wang2020attack,xie2019dba,chen2017targeted,liu2017trojaning,nguyen2020poisoning,gu2017badnets, li2021invisible}. 
The backdoor attack is more difficult to be detected compared with the untargeted poisoning attacks \cite{bhagoji2019analyzing, biggio2012poisoning,liu2017trojaning,chen2017targeted,gu2017badnets} since it does not affect the regular function of the model and its gradient is more similar to benign ones \cite{wang2020attack, nguyen2022flame}. Multiple defenses have been proposed to improve the robustness of FL, such as the scoring-based methods \cite{fung2020limitations,blanchard2017machine,yin2018byzantine,pillutla2022robust,cao2020fltrust,DBLP:conf/aaai/ZhaoSWJ22,fu2019attack,li2020learning,andreina2021baffle,guerraoui2018hidden}, which leverage a particular metric to distinguish malicious gradients from the benign ones. Despite its effectiveness against some backdoors, researchers discovered that well-designed attacks (termed \textit{stealthy backdoors}) whose gradients are indistinguishable from benign ones (through scaling \cite{bagdasaryan2020backdoor, wang2020attack} or trigger split \cite{xie2019dba}) could easily bypass these defenses. Differential privacy (DP) -based method \cite{sun2019can, ozdayi2021defending, nguyen2022flame,xie2021crfl} builds upon the observation that the DP method \cite{dwork2014algorithmic}, traditionally used against DP attacks, is also effective against backdoors. By adding Gaussian noise to the global model, these methods can dilute the impact of potentially poisoned model updates. Surprisingly, DP-based methods show great ability in resisting \textit{stealthy backdoors} (\eg, Edge-case PGD \cite{wang2020attack}).
Despite its ability to resist the \textit{stealthy backdoors}, the noises added by the DP significantly decrease the overall performance and the convergence speed. In comparison, the distance-based method less impacts the global model by aggregating only the benign gradients. Consequently, a natural question arises: \textit{can we defend the stealthy backdoors without sacrificing the performance of the FL model?} To accomplish this, we turn to the distance-based methods that don't sacrifice the benign performance and promote the research question: \textit{how can we successfully leverage distance metrics to discriminate hostile updates from benign ones?}

To this end, we revisit the distance-based defense and discover two limitations: 1. Euclidean distance (\ie, $L_2$ distance) suffers from the curse of dimensionality. Parameters of Neural Networks (NNs) can be viewed as high-dimensional vectors, and Euclidean distance generally fails to discriminate between malicious and benign gradients in high-dimensional space. 2. Single metric takes effect only against particular attacks with detailed assumptions regarding the malicious gradients. For instance, cosine distance detects malicious gradients with sizeable angular deviations while euclidean distance detects malicious gradients with a large $L_2$ norm scaled by the attacker to impact the global models. Moreover, backdoor attacks are conducted with different data and scenarios, resulting in malicious gradients with various characteristics that a single metric cannot handle (\ie, both gradients with a large norm and gradients with sizeable angular deviations exist in one round of aggregation). What's worse is that the defender has no knowledge regarding the attacker and the underlying data distributions due to privacy requirements by FL, which makes detecting with a single metric even more difficult. 

To address the above two problems, we first introduce the Manhattan distance, which we theoretically prove more meaningful in high dimensional space than Euclidean distance. Empirically, with the Manhattan distance, our proposed distance-based defense shows remarkable performance against the \textit{stealthy backdoors}. To cope with gradients with various properties, we leverage multiple metrics cooperatively to identify the malicious gradients. We demonstrate in Section \ref{ablation study} and Figure \ref{weight} that malicious gradients of some characteristics are better identified by specific metrics, which justify our motivation. To handle the different attacks and environments, we further propose to apply a whitening transformation and generate dynamic weights to handle the non-IID distribution of participants and different scales brought by the different distances. 
Finally, we compute the score for each submitted gradient and aggregate only the benign ones based on the scoring.
Extensive experiments illustrate that our defense maintains a high model performance and robustness simultaneously, which has never been achieved before. At a high level, we summarize the main contributions of this paper as follows:
\begin{itemize}
    \item We present a novel defense with multi-metrics to adaptively identify backdoors, which is applicable in a generic adversary model without predefined assumptions over the attack strategy or data distribution.
    \item We show that by introducing the Manhattan distance, our defense alleviates the “meaningfulness” problem of Euclidean distance in high dimensions. By utilizing multiple metrics with dynamic weighting, our defense can resist backdoor attacks under a wide range of attack settings and data distributions.
    \item We empirically evaluate our method on various datasets and different attack settings. By showing that our defense maintains both high robustness and benign performance under the \textit{stealthy backdoors} that previously have not been successfully defended, we demonstrate the effectiveness of our approach.
\end{itemize}

\section{Related work}
\textbf{Backdoor Attacks on Federated Learning.}
FL enables multiple clients collaboratively train a model without local data exchange. In an FL system, clients update the local model with private data and submit the parameters to the server for aggregation. Given a total of $N$ participants, each with $n^{(i)}$ samples. In each round $t$, the server randomly selects $K$ clients to participate. FedAvg \cite{mcmahan2017communication} is a widely-adopted FL algorithm, which can be formulated as: 
\begin{equation}
\boldsymbol{w}_{t}=\boldsymbol{w}_{t-1}+\eta \cdot \frac{\sum_{i=1}^{K} n^{(i)} \Delta w_{t}^{(i)}}{\sum_{i=1}^{K} n^{(i)}}
\end{equation}
where $w_t$ is the global model parameters at round $t$, $\Delta w_t^{(i)}$ is the local model updates from the client $i$ in round $t$, and $\eta$ is the global model learning rate. 

Empirical studies from Bagdasaryan et al. \cite{bagdasaryan2020backdoor} demonstrate that FedAvg \cite{mcmahan2017communication} is vulnerable to backdoor attacks since FL has no authority over local data and the training process due to privacy concerns. The model replacement attack \cite {bagdasaryan2020backdoor} successfully injects a backdoor into the global model by the single-shot attack. Some well-designed attacks strategically target the weakness of the defenses, such as the PGD attack \cite{wang2020attack} scaling and projecting the gradients, or the DBA attack \cite{xie2019dba} splitting the triggers before uploading them. Moreover, the Edge-case PGD attack \cite{wang2020attack} modifies the poisoned data and model. Undoubtedly, these attacks present a tremendous challenge to the security of the FL system.

\textbf{Defense Against the Backdoor Attack.}
Various methods to defend against backdoor attacks improve the robustness and security of FL. In general, we can divide these methods into two categories. 
The first type of defense strategies tries to distinguish malicious gradients from benign ones by classifying or scoring with the assumption that benign and malicious gradients show distinct differences in the vector space \cite{fung2020limitations,blanchard2017machine,yin2018byzantine,pillutla2022robust,cao2020fltrust,DBLP:conf/aaai/ZhaoSWJ22,fu2019attack,li2020learning,andreina2021baffle,guerraoui2018hidden}. Krum and Multi-Krum \cite{blanchard2017machine} select the gradient with the smallest Euclidean norm from the others to aggregate and update the global model at each round. Foolsgold \cite{fung2020limitations} assumes that the gradients of attackers are always consistent and distinguishes them by the Cosine similarity between historical updates. Trimmed Mean and Median \cite{yin2018byzantine} remove a fraction of the gradients with maximum and minimum values and then take the median value on each dimension of the remaining gradients as the parameters of the global model. RFA \cite{pillutla2022robust} leverages the geometric median of the gradients to keep the aggregated model close to the benign model and away from the backdoor one. The strength lies in that, if successful, the gradients of attackers are excluded from the aggregation process, which solves the core and fundamental problem of defending the attack.

However, the above methods take effect only under particular attacks since they make detailed assumptions about the attack or the data distributions. Foolsgold \cite{fung2020limitations} assumes the gradients of attackers are consistent while benign gradients from Non-IID data differ. Krum \cite{blanchard2017machine} assumes that the data of benign clients is similar under IID settings and excludes abnormal malicious ones. These predefined assumptions limit the applicability of these defenses under various attacks and data distributions and can be easily bypassed by elaborate attacks. For example, PGD \cite{wang2020attack} breaks any Euclidean-based defense (\eg Krum and Multi-Krum) by scaling the norm to a small value. Other methods, such as FLTrust \cite{cao2020fltrust}, use a root dataset and aggregate only those gradients with large cosine similarities with the root dataset. FedInv \cite{DBLP:conf/aaai/ZhaoSWJ22} inverts the gradients into dummy datasets and then identifies the malicious gradient by comparing these dummy datasets. These methods are limited as well. FLTrust cannot defend against backdoors crafted on edge cases( \eg Edge-case PGD) since they appear less often and are probably not in the root dataset. Inverting the gradients into dummy datasets violates the privacy requirement of FL. Essentially, this method can be characterized as another type of attack \cite{augenstein2019generative, hao2019towards}. 

The other category of defenses \cite{sun2019can, ozdayi2021defending, nguyen2022flame,xie2021crfl} builds upon the observation that the differential privacy (DP) method \cite{dwork2014algorithmic}, traditionally used against DP attacks is also effective against backdoor attacks. Earlier studies \cite{bagdasaryan2020backdoor} have demonstrated that clipping and adding noise (typically used in DP) can resist backdoor attacks without any specific assumptions over the data distribution. Following these findings, Weak-DP \cite{sun2019can} clips the norm and adds the Gaussian noise to the global model for mitigating the backdoors. RLR \cite{ozdayi2021defending} and Flame \cite{nguyen2022flame} apply the clustering methods before the DP process. However, the noise added significantly decreases the performance and the convergence speed of the FL system. In a nutshell, using DP sacrifices regular functionality for robustness. 
Despite its deficiency, the DP-based method has been the only method that has taken effect against \textit{stealthy backdoors} \cite{nguyen2022flame}.

\begin{figure*}[!t]
\centering
\includegraphics[width=0.928\textwidth]{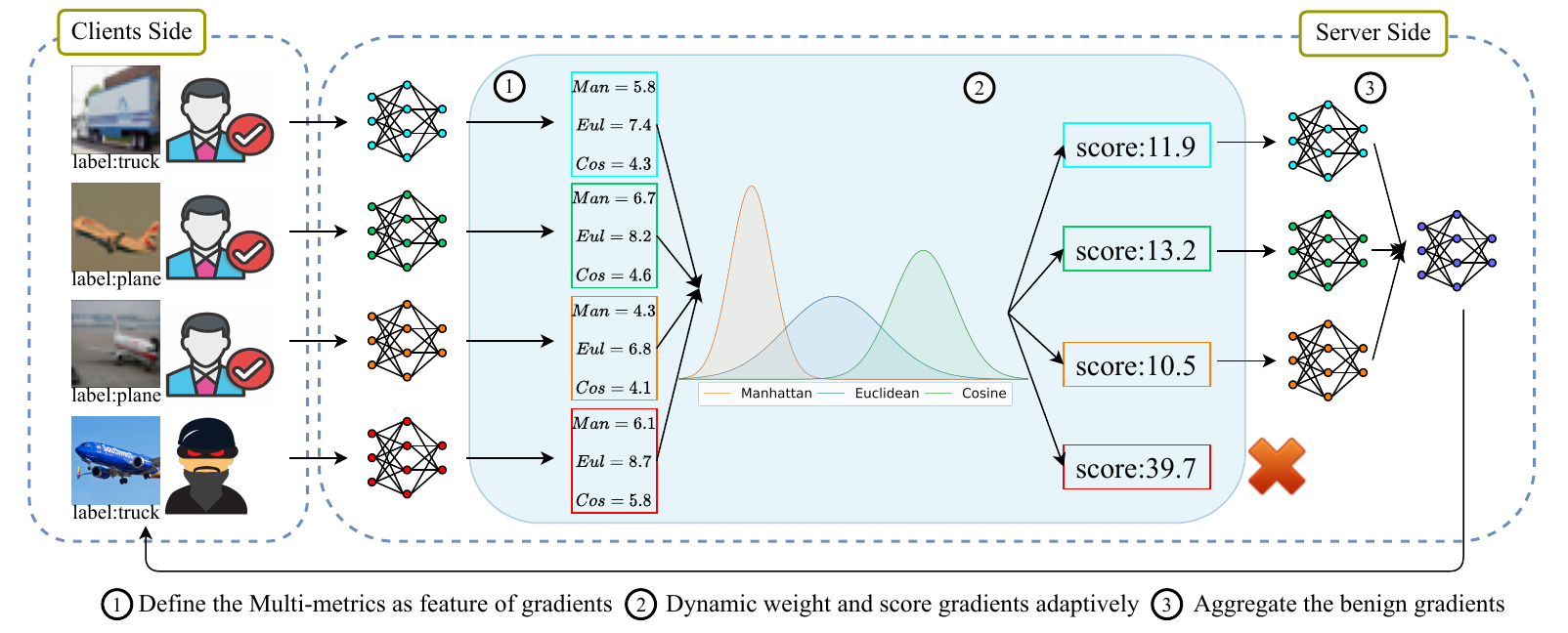}
\caption{The overview of our defense process. Step 1 and step 2 are our core contributions.}
\label{overview}
\end{figure*}

\section{Methodology}

We first provide the threat model in Section \ref{sec:threat} and then detail our design of the proposed defense method. Our goal is to design an efficient defense with the following properties: (1) Effectiveness: To prevent the adversary from achieving its hostile purpose, we aim to design a defense that identifies as many malicious updates as possible to eliminate its impact on the global model. (2) Performance: Benign performance of the global model must be preserved to maintain its utility. (3) Independence: The defense must be applicable to generic adversary models under different data distributions and attack strategies. 
Our defense consists of three major components: defining and calculating the gradient feature, computing the dynamic weight and score for each update, and aggregating the benign gradients with the score computed, as shown in Figure \ref{overview} and Algorithm \ref{algorithm}.

\begin{figure}[htbp]
\centering
\includegraphics[width=2in]{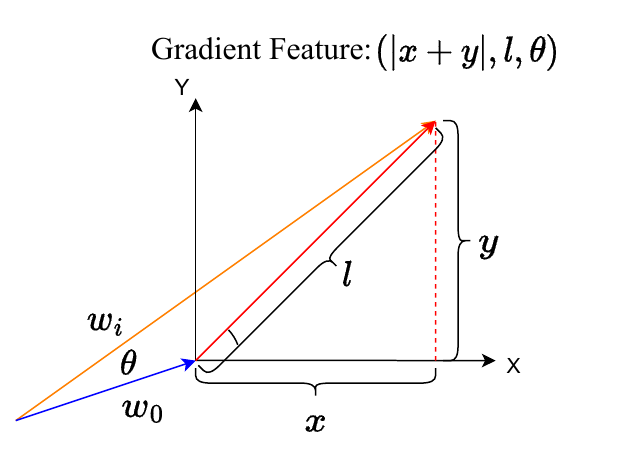}
\caption{Demonstration of  \textit{feature of gradient} on two dimensional space. $\boldsymbol{w}_0$ is initial global model $\boldsymbol{w}_0$ and $\boldsymbol{w}_i$ is the local model of the $i$-th client. $\boldsymbol{w}_i - \boldsymbol{w}_0$ is the gradient and we define the \textit{feature of gradient} as $(|x+y|,l,\theta)$.}
\label{gradient_feature}
\end{figure}

\subsection{Threat Model}
\label{sec:threat}
\textbf{Adversarial goals}. 
Adversary aims to alter the behavior of the model on some specific examples with no negative impact on the benign performance. We consider the adversary effective if this is met. In addition, the adversary will make their attack as covert as possible, \eg, by scaling down. 
Formally, adversary $\mathcal{A}$ manipulates the global model $G$ to the poisoned model $G'$. For any inputs $x$, we have the outputs below:
\begin{equation}
    G'\left(x\right)=\left\{\begin{array}{cc}
    l^{\prime} \neq G(x) & \forall x \in D_{\mathcal{A}} \\
    G(x) & \forall x \notin D_{\mathcal{A}}
\end{array}\right.
\end{equation}
where $D_{\mathcal{A}}$ denotes the poisoned dataset belonging to adversary $\mathcal{A}$, and $l'$ denotes the specific label modified by adversary $\mathcal{A}$ of the input $x$. 

\textbf{Adversarial capabilities}. This paper assumes the white-box setting with full knowledge of aggregation operations, including potential defenses. Also, adversaries can freely alter the parameters in the local training, ensure participation in each round and compromise less than 50\% of total FL clients. However, the adversary cannot control any process or parameter executed in the aggregator.

\subsection{Feature of gradient}
The essential logic of distance-based defense involves defining some indicative metric that can well discriminate the malicious gradients from the benign ones and removes the hostile updates from the aggregation. Consequently, the core problem becomes how to define a metric that identifies the characteristics of hostile gradients. For instance, NDC \cite{sun2019can} utilizes the length of the gradient (\ie  Euclidean distance) while FoolGold \cite{fung2020limitations} leverages Cosine similarity to measure the difference between updates. 

\textbf{Manhattan Distance.}
However, the Euclidean distance suffers from the so-called “curse of dimensionality” \cite{wright1962adaptive}, which renders distance metrics less sensitive in high dimensional space, especially in the case of Euclidean distance. Theorem \ref{theo} uses $\frac{Dmax_d^k-Dmin_d^k}{D\min_d^k}$ to indicate the “meaningfulness” of a distance referred as the \textit{Relative Contrast}. It demonstrates that the distance metric is meaningless in high-dimensional space because as dimensionality increases, the distance between the nearest and the furthest neighbor approximates to be the same.
\begin{theorem1}[Beyer et. al.\cite{beyer1999nearest} The curse of dimensionality]
\label{theo}
\begin{equation}
\text{If } \lim _{d \rightarrow \infty} var\left(\frac{\left\|X_d\right\|_k}{E\left[\left\|X_d\right\|_k\right]}\right)=0,
\text{then }\notag
\end{equation}
\begin{equation}
\frac{Dmax_d^k-Dmin_d^k}{D\min_d^k} \rightarrow_p 0, \notag
\end{equation}
where $d$ denotes the dimensionality of the data space, $E\left[X\right]$ and $var\left[X\right]$ denotes the expected value and variance of a random variable $X$, $Dmax_d^k$ and $Dmin_d^k$ means the farthest/nearest distance of the $N$ points to the origin using the distance metric $L_k$.
\end{theorem1}

To find out which distance metric is more meaningful in high-dimensional space, we size up the relation between the dimension $d$ and the distance $L_k$. 
We have Lemma \ref{lem}, which shows that $Dmax_d^k-Dmin_d^k$ increases at the ratio of $d^{(1 / k)-(1 / 2)}$.
\begin{lemma1}[Hinneburg et.al.\cite{hinneburg2000nearest}]
\label{lem}
Let $\mathcal{F}$ be an arbitrary distribution of two points and the distance function $\|\cdot\|$ be an $L_k$ metric. Then,
\begin{equation}
\lim _{d \rightarrow \infty} E\left[\frac{Dmax_d^k-Dmin_d^k}{d^{(1 / k)-(1 / 2)}}\right]=C_k, \notag
\end{equation}
where $C_k$ is some constant dependent on $k$.
\end{lemma1}
Subsequently, we use the value of $Dmax_d^k-Dmin_d^k$ as a criterion to evaluate the ability of a particular metric to discriminate different vectors in high dimensional space.
\begin{proposition1}
\label{prop}
Let $M_d=Dmax_d^1-Dmin_d^1$ reflect the discriminating ability of Manhattan distance and $U_d=Dmax_d^2-Dmin_d^2$ reflect the discriminating ability of Euclidean distance. Then, 
\begin{equation}
\lim _{d \rightarrow \infty} E\left[\frac{M_d}{U_d \cdot d^{\frac{1}{2}}}\right] = C'. \notag
\end{equation}
where $C'$ is a constant. $Proof$ see in Appendix \ref{proof_prop}.
\end{proposition1}
Proposition \ref{prop} shows that $M_d$ and ${U_d \cdot d^{\frac{1}{2}}}$ are infinite of the same order. Since $d^{\frac{1}{2}}$ is also an infinite value, $M_d$ towers above $U_d$, suggesting that Manhattan can discriminate more than the Euclidean distance in high-dimension space. Given this, we conclude that the Manhattan distance is better than the Euclidean metric, withstanding the curse of dimensionality, which chimes in with the previous experiment \cite{aggarwal2001surprising,mirkes2020fractional}. The parameters of neural networks can be viewed as a typical high-dimensional space, especially when the number of parameters increases as the trend in deep learning applications \cite{brown2020language, shao2021intern, ramesh2022hierarchical, fedus2021switch, liu2022swin, yu2022coca, riquelme2021scaling}. In light of this and Theorem \ref{theo}, we argue that Euclidean distance is insufficient to discriminate between malicious and benign gradients. With  Proposition \ref{prop}, we introduce the Manhattan distance metric to describe the characteristics of gradients which captures the difference in high dimensional space better.

\textbf{Multiple Metrics as Feature of Gradient.}
Another problem that we identify as discussed in Section \ref{sec:intro} is that current methods conduct defense on a single metric basis. With only one metric, a sophisticated attacker could effortlessly bypass them with a well-designed gradient. For instance, PGD attack \cite{wang2020attack} manage to break the defense based on Euclidean distances by clipping the gradient norm to a small value. Moreover, attackers conduct attacks under different environments and data distributions, leading to malicious gradients with diverse characteristics that a single metric cannot handle. 
To this end, we propose multiple metrics to cooperatively identify the malicious gradient by defining the \textit{feature of gradient} as the angle by the Cosine similarity, length by the Euclidean distances, and its Manhattan norm by the Manhattan distance, namely, $\boldsymbol{x} = (x_{Man}, x_{Eul}, x_{Cosine})$. Here, $x_{Man}$ denotes the Manhattan distance, $x_{Eul}$ denotes the Euclidean distance and $x_{Cosine}$ denotes the Cosine similarity. 
Figure \ref{gradient_feature} provides a schematic diagram of the \textit{gradient feature}. We compute the gradient feature $x_{Man}$, $x_{Eul}$, and $x_{Cosine}$ of the $i$-th client of each round as follows:
$x_{Man}^{(i)}={\left\|\boldsymbol{w}_{i}-\boldsymbol{w}_{0}\right\|_{1}}$, $x_{Eul}^{(i)}={\left\|\boldsymbol{w}_{i}-\boldsymbol{w}_{0}\right\|_{2}}$, $x_{Cosine}^{(i)}=\frac{\left\langle\boldsymbol{w}_{0} \boldsymbol{w}_{i}\right\rangle}{\left\|\boldsymbol{w}_{0}\right\| \cdot\left\|\boldsymbol{w}_{i}\right\|}$, where $\boldsymbol{w}_0$ denotes the global model before federated training and $\boldsymbol{w}_i$ denotes the $i$-th client's local model after training, $\left\|\cdot\right\|_1$ means the $L_1$ norm of a vector, $\left\|\cdot\right\|_2$ means the $L_2$ norm of a vector, $\left\langle\cdot, \cdot\right\rangle$ and represents inner product. After defining the \textit{feature of gradient}, we utilize it for malicious identification. The goal is to identify the outlier among the gradients. 
We use the sum of the distance between each gradient as the indicator.
Formally, we define it as Equation \ref{redefinition}.

\begin{equation}
    \label{redefinition}
    \begin{split}
        &\boldsymbol{x}'^{(i)}=(\sum_{j,j \neq i}^{K}\lvert{x_{Man}^{(i)}-x^{(j)}_{Man}\rvert}, \hfill  
        \\&\sum_{j,j \neq i}^{K}\lvert{x_{Eul}^{(i)}-x^{(j)}_{Eul}\rvert}, \sum_{j,j \neq i}^{K}\lvert{x_{Cosine}^{(i)}-x^{(j)}_{Cosine}\rvert}),\hfill
    \end{split}
\end{equation}

\subsection{Dynamic Weighting Through Whitening}
\label{weighting_section}
After defining the \textit{feature of gradient} and indicator, we now provide an approach to score each gradient. Two factors are taken into account when considering an efficient way to turn the indicator into a unified score that works under various data distributions and attacks.
\begin{itemize}
    \item Different scales of the three distance metrics are the first obstacle before utilizing these metrics collaboratively. Since each metric is correlated, a novel regularization is required instead of the usual normalization by the maximum value.
    \item Different data distributions (\eg different degrees of non-IID) render the gradients of both malicious and benign clients different. Thus, dynamic weighting is required to cope with various environments and attacks to achieve a universal defense.
\end{itemize}

Considering the above factors, we propose first to project the value of the metrics in the \textit{feature of gradient} to its corresponding principal axis by applying a whitening process:

\begin{equation}
\delta^{(i)}=\sqrt{\boldsymbol{x'}^{(i)\top} \boldsymbol{\Sigma}^{-1}\boldsymbol{x'}^{(i)}}.
\end{equation}
where $\boldsymbol{\Sigma}$ is the the covariance matrix of $\boldsymbol{X}=[\boldsymbol{x}'^{(1)} \boldsymbol{x}'^{(2)}\cdots \boldsymbol{x}'^{(K)}]^\top$. Since we need to calculate the inverse of the covariance matrix $\boldsymbol{\Sigma}$, the number of samples must be larger than the number of features, i.e., $K > 3$. Please notice that the inverse of the covariance matrix is calculated based on the selected gradients, which will change the weights of the features dynamically depending on the feature distribution, hence "dynamic weighting". With such a dynamic weight, our method can better adapt to different environments and resist various attacks.

\subsection{Aggregation of Benign Gradients}
\label{p_section}
After obtaining the score of each gradient, we aggregate the gradients with high scoring. A higher score indicates that the gradient is less divergent among all. After the selection of "benign" gradients, we can aggregate them with any existing aggregation methods (e.g. Averaging of gradients). We follow \cite{mcmahan2017communication} and perform standard FedAvg on the "benign" gradients. A detailed analysis of aggregation of benign gradients is in  
Appendix \ref{p_ablation}.

\begin{algorithm} 
    \floatname{algorithm}{Algorithm}
    \caption{Our aggregation algorithm} 
    \label{algorithm}
    \hspace*{0.02in} {\bf Input:} Total number of clients in each round $K$, models of clients $\boldsymbol{w}^{(1)}$,$\boldsymbol{w}^{(2)},...,\boldsymbol{w}^{(K)}$, last global model $\boldsymbol{w}_{0}$,fraction selected to aggregation $p$, global learning rate $\eta$, size of the $i-th$ client dataset $n^{(i)}$\\
    \hspace*{0.02in} {\bf Output:}  Global model $w^{*}$
    \begin{algorithmic}[1]
        \For{$i \in \{ 1, 2,  ..., K \}$} \Comment{compute the gradients features}
            \State $x_{Man}^{(i)} \gets {\left\|\boldsymbol{w}_{0}-\boldsymbol{w}_{i}\right\|_{1}}$
            \State $x_{Eul}^{(i)} \gets {\left\|\boldsymbol{w}_{0}-\boldsymbol{w}_{i}\right\|_{2}}$
            \State $x_{Cos}^{(i)} \gets \frac{\left\langle\boldsymbol{w}_{0},\boldsymbol{w}_{i}\right\rangle}{\left\|\boldsymbol{w}_{0}\right\| \cdot\left\|\boldsymbol{w}_{i}\right\|}$
            \State $\boldsymbol{x}^{(i)} \gets (x_{Man}^{(i)},x_{Eul}^{(i)},x_{Cos}^{(i)})$ 
        \EndFor
        
        \For{$i,j \in \{ 1, 2,  ..., K \}$} \Comment{compute the sum of the distance be-
tween each gradient}
            \State ${x'}^{(i)}_{Man} \gets \sum_{j,j \neq i}^{K}\lvert{x_{Man}^{(i)}-x^{(j)}_{Man}\rvert}$  
            \State ${x'}^{(i)}_{Eul} \gets \sum_{j,j \neq i}^{K}\lvert{x_{Eul}^{(i)}-x^{(j)}_{Eul}\rvert}$
            \State ${x'}^{(i)}_{Cos} \gets \sum_{j,j \neq i}^{K}\lvert{x_{Cos}^{(i)}-x^{(j)}_{Cos}\rvert}$
            \State $\boldsymbol{x'}^{(i)} \gets ({x'}_{Man}^{(i)},{x'}_{Eul}^{(i)},{x'}_{Cos}^{(i)})$ 
        \EndFor
        \State $X \gets [\boldsymbol{x'}^{(1)},\boldsymbol{x'}^{(2)},...,\boldsymbol{x'}^{(K)}]^\top$
        \State $\Sigma \gets$ the Covariance of $X$ \Comment{compute the adaptive weight matrix in this round}
        \For{$i \in \{ 1, 2,  ..., K \}$}
            \State $\delta^{(i)} \gets \sqrt{\boldsymbol{x'}^{(i)\top} \cdot \Sigma^{-1} \cdot \boldsymbol{x'}^{(i)}}$ \Comment{dynamically compute the divergency}
        \EndFor 
        \State Remove the $K\cdot (1-p)$ models with the high divergency $\delta^{(i)}$, remaining models as $\mathbb{B}$ 
        \State $\boldsymbol{w}^{*} \gets \boldsymbol{w}_{0}+\eta \cdot \frac{\sum_{i\in \mathbb{B}} n^{(i)}\cdot (\boldsymbol{w}^{(i)}-\boldsymbol{w}_{0})}{\sum_{i\in \mathbb{B}} n^{(i)}}$ 
    \end{algorithmic} 

\end{algorithm}
\section{Experiment Setup}
We will first introduce the settings and then the objective of our experiments.

\textbf{Datasets.}
We conduct our experiments on multiple datasets, including two visual datasets, CIFAR-10 \cite {krizhevsky2009learning} and EMNIST \cite{cohen2017emnist}. We also experiment on other datasets to prove the generalization ability, including CINIC10 \cite{darlow2018cinic}, LOAN \cite{loan} and Sentiment140 \cite{go2009twitter} as follows:
\begin{itemize}
\setlength{\itemsep}{0pt}
\setlength{\parsep}{0pt}
\setlength{\parskip}{0pt}
    \item \textbf{CIFAR-10 dataset \cite{krizhevsky2009learning}} is a vision classification dataset containing $50,000$ training samples and $10,000$ testing samples with ten classes. To simulate the FL environment, we set the $a$ of the Dirichlet distribution to be $0.5$, representing a non-IID distribution. Smaller $a$ means a larger degree of non-IID. 
    \item \textbf{EMNIST dataset \cite{cohen2017emnist}} is a digit classification dataset consisting of $280,000$ real-world handwritten images of digits from $0$ to $9$. We also set the degree of non-IID to be $a=0.5$.
    \item \textbf{CINIC10 \cite{darlow2018cinic}} has a total of 270,000 images, 4.5 times that of CIFAR-10, which is constructed from two different sources: ImageNet and CIFAR-10. We set the non-IID degree as $a=0.5$.
    \item \textbf{Lending Club Loan dataset (LOAN) \cite{loan}} contains financial information such as loan information and credit scores for loan status prediction. There are 2,260,668 samples with a total of 9 classes. To simulate a more realistic situation, we divide the data by US states, where each state represents one client. We set the non-IID parameter $a = 0.9$. 
    \item \textbf{Sentiment140 \cite{go2009twitter}} is a publicly available English Twitter sentiment classification dataset containing 1.6 million tweets, of which 800,000 are for training, and the other 800,000 are for testing. Each tweet in the dataset is labeled with a sentiment polarity (positive, neutral, or negative). 
\end{itemize}


\textbf{Models.} 
Different models are used on different datasets to demonstrate the generality of our approach. A 9-layer VGG style network (VGG-9) \cite{simonyan2014very} is trained on the CIFAR-10. On the EMNIST, we train LeNet-5 model \cite{lecun1998gradient}. We train the LSTM model \cite{hochreiter1997long} for Sentiment140 and a network with three fully-connected layers for LOAN.

\textbf{Backdoor Attacks.}
We implement both pixel-pattern and semantic backdoors. To make the attack more challenging, we use the multiple-shot attack instead of the single-shot. 
The following are the attack methods we use:
\begin{itemize}
\setlength{\itemsep}{0pt}
\setlength{\parsep}{0pt}
\setlength{\parskip}{0pt}
\item \textbf{Model Replacement Attack} \cite{bagdasaryan2020backdoor}: Malicious clients generate backdoor gradients during their local training. Attackers scale up their gradients to ensure that the malicious model will replace the global model. Usually, the scaling factor is $N/K$.;
\item \textbf{DBA Attack} \cite{xie2019dba}: DBA decomposes a global trigger pattern into separate local patterns and embeds them into the training sets of different attackers, resulting in a more negligible difference between benign and backdoor gradients;
\item \textbf{PGD Attack} \cite{wang2020attack}: To bypass the Euclidean-based defense, PGD Attack projects the model parameter within the ball centered around the global model; 
\item \textbf{Edge-case PGD Attack} \cite{wang2020attack}: Attackers have a set of edge-case samples and some benign samples as their poisoned datasets. Edge-case samples are unlikely to exist in the benign training set. Thus, the benign gradients cannot dilute the backdoor.
We launch this attack in the form of PGD. The Edge-case PGD attack is so stealthy that most defenses cannot defend against it. 
\end{itemize}


\textbf{Evaluation.}
We leverage the following two metrics for evaluating the effectiveness of our defense:
\begin{itemize}
\setlength{\itemsep}{0pt}
\setlength{\parsep}{0pt}
\setlength{\parskip}{0pt}
\item \textbf{Backdoor Accuracy (BA)} indicates whether the attacker succeeds. It shows the accuracy of the model on the backdoor task. The attackers aim to maximize this metric, and the lower the BA, the more efficient the defense.
\item \textbf{Main Task Accuracy (MA)} is the accuracy on the main task. The overall objective of all participants in an FL system is to maximize MA. The benign clients maximize the model performance, while the attackers seek to be stealthy and maintain MA. Thus, the defense should not cause a significant drop in MA, which we report to demonstrate that our defense does not affect the functionality of the FL system. 
We report all BA and MA values in percentages. 
\end{itemize}

\textbf{Hyperparameters} are detailed in Appendix \ref{hyperparameters}.

\textbf{Objectives of Experiments.} Firstly, we aim to illustrate the effectiveness of our proposed defense under different attacks by comparing it with the current and previous state-of-the-art defenses.
Secondly, we demonstrate that the performance of our defense is consistent under different scenarios (\eg, different degrees of non-IID $a$ and different attack frequencies). Thirdly, we figure out the reason for our defense works through the ablation study in Section \ref{ablation study}.
\section{Experiment Results}

\begin{table*}[htbp]
\caption{Robustness of our approach compared to the SOTA defenses for various challenging attacks.}
\label{compare_table}
\centering
\resizebox{1\textwidth}{!}{%
$\begin{tabular}{@{}clccccccccc@{}}
\toprule
\multirow{2}{*}{Dataset} & \multicolumn{1}{c}{\multirow{2}{*}{Defense}}            & \multicolumn{2}{c}{Model Replacement \cite{bagdasaryan2020backdoor}} & \multicolumn{2}{c}{DBA \cite{xie2019dba}} & \multicolumn{2}{c}{PGD \cite{wang2020attack}} & \multicolumn{2}{c}{Edge-case PGD \cite{wang2020attack}} & \multirow{2}{*}{Ranking Score $\uparrow$} \\ \cmidrule(lr){3-10}
                         & \multicolumn{1}{c}{}                                    & MA $\uparrow$               & BA $\downarrow$               & MA $\uparrow$  & BA $\downarrow$ & MA $\uparrow$    & BA $\downarrow$   & MA $\uparrow$         & BA $\downarrow$        &                                                     \\ \midrule
\multirow{8}{*}{\rotatebox{90}{CIFAR10}} & FedAvg \cite{mcmahan2017communication} & \textbf{86.95}                                    & 64.80                                      & 79.23                       & 90.44                        & 87.04                         & 14.44                          & 87.14                              & 55.10                               & 0                                                   \\
                         & RFA \cite{pillutla2022robust}          & 86.69(+0.00)                             & 25.56(-0.61)                               & 79.6(+0.00)                 & 57.69(-0.36)                 & \textbf{87.1}(+0.00)                   & 52.56(+2.64)                   & 86.47(-0.01)                       & 65.31(+0.19)                        & -1.86                                               \\
                         & Foolsgold \cite{fung2020limitations}   & 85.71(-0.01)                             & 6.67(-0.90)                                & 77.56(-0.02)                & 3.43(-0.96)                  & 84.92(-0.02)                  & 14.44(+0.00)                   & 85.72(-0.02)                       & 45.41(-0.18)                        & +1.96                                               \\
                         & Krum \cite{blanchard2017machine}       & 82.17(-0.05)                             & 6.11(-0.91)                                & 78.18(-0.01)                & 6.01(-0.93)                  & 82.32(-0.05)                  & 66.67(+3.62)                   & 81.23(-0.07)                       & 59.18(+0.07)                        & -2.04                                               \\
                         & Multi-Krum \cite{blanchard2017machine} & 86.55(+0.00)                             & 1.67(-0.97)                                & 79.33(0.00)                 & 91.39(+0.01)                 & 86.52(-0.01)                  & 17.78(+0.23)                   & \textbf{87.4}(+0.00)                        & 60.2(+0.09)                         & +0.63                                               \\
                         & Weak-DP \cite{sun2019can}              & 74.41(-0.14)                             & 46.11(-0.29)                               & 10.00(-0.87)                & \textbf{0.00}(-1.00)                  & 73.61(-0.15)                  & 12.78(-0.11)                   & 73.84(-0.15)                       & 53.06(-0.04)                        & +0.12                                               \\
                         & Flame \cite{nguyen2022flame}           & 80.58(-0.07)                             & \textbf{0.56}(-0.99)                                & 76.78(-0.03)                & 37.24(-0.59)                 & 81.24(-0.07)                  & \textbf{0.56}(-0.96)                    & 81.41(-0.07)                       & 5.12(-0.91)                         & +3.21                                               \\ \cmidrule(l){2-11} 
                         & Ours                                                    & 86.34(-0.01)                             & \textbf{0.56}(-0.99)                                & \textbf{79.61}(+0.00)                & 9.98(-0.89)                  & 86.44(-0.01)                  & 0.56(-0.96)                    & 86.86(+0.00)                       & \textbf{3.06}(-0.94)                         & \textbf{+3.77}                                               \\ \midrule
\multirow{8}{*}{\rotatebox{90}{EMNIST}}  & FedAvg \cite{mcmahan2017communication} & 99.54                                    & 96.00                                      & 97.68                       & 94.13                        & \textbf{99.55}                         & 10.00                          & 99.37                              & 96.00                               & 0                                                   \\
                         & RFA \cite{pillutla2022robust}          & 99.57(+0.00)                             & 6.00(-0.94)                                & \textbf{97.87}(+0.00)                & 1.39(-0.99)                  & 99.32(+0.00)                  & 4.00(-0.60)                    & 99.29(+0.00)                       & 97.00(+0.01)                        & +2.51                                               \\
                         & Foolsgold \cite{fung2020limitations}   & 96.42(-0.03)                             & 98.00(+0.02)                               & 97.24(+0.00)                & 0.64(-0.99)                  & 99.07(+0.00)                  & 94.00(+8.40)                   & 99.13(+0.00)                       & 98.00(+0.02)                        & -7.49                                               \\
                         & Krum \cite{blanchard2017machine}       & 99.22(+0.00)                             & \textbf{0.00}(-1.00)                                & 97.7(+0.00)                 & 0.56(-0.99)                  & 99.12(+0.00)                  & 1.00(-0.90)                    & 99.14(+0.00)                       & 12.00(-0.88)                        & +3.76                                               \\
                         & Multi-Krum \cite{blanchard2017machine} & \textbf{99.58}(+0.00)                             & \textbf{0.00}(-1.00)                                & 97.85(+0.00)                & 47.43(-0.50)                 & 99.54(+0.00)                  & 0.00(-1.00)                    & 99.57(+0.00)                       & 84.00(-0.13)                        & +2.63                                               \\
                         & Weak-DP \cite{sun2019can}              & 99.37(+0.00)                             & 86.00(-0.10)                               & 10.00(-0.90)                & \textbf{0.00}(-1.00)                  & 99.41(+0.00)                  & 14.00(+0.40)                   & 99.39(+0.00)                       & 89.00(-0.07)                        & -0.12                                               \\
                         & Flame \cite{nguyen2022flame}           & 99.39(+0.00)                             & \textbf{0.00}(-1.00)                                & 97.12(-0.01)                & 17.38(-0.82)                 & 99.39(+0.00)                  & \textbf{0.00}(-1.00)                    & 99.44(+0.00)                       & 13.00(-0.86)                        & +3.67                                               \\ \cmidrule(l){2-11} 
                         & Ours                                                    & 99.53(+0.00)                             & \textbf{0.00}(-1.00)                                & 97.39(+0.00)                & 4.23(-0.96)                  & 99.54(+0.00)                  & \textbf{0.00}(-1.00)                    & \textbf{99.58}(+0.00)                       & \textbf{0.00}(-1.00)                         & \textbf{+3.95}                                               \\ \bottomrule
\end{tabular}%
$}
\end{table*}

\subsection{Robustness against Different Attacks and Comparison with SOTA}
\label{compare_defenses}
We first demonstrate the robustness of our approach under different attacks and then compare our methods with the state-of-the-art defense methods such as Krum \cite{blanchard2017machine}, Multi-Krum \cite{blanchard2017machine}, Foolsgold \cite{fung2020limitations}, RFA \cite{pillutla2022robust}, Weak-DP \cite{sun2019can}, and Flame \cite{nguyen2022flame} in Table \ref{compare_table}.
All the defenses except RFA and Weak-DP successfully defended against the model replacement attack since RFA and Weak-DP still incorporate some malicious gradient into the global model. Our defense achieves the lowest BA on both datasets($0.56\%$ and $0.00\%$) with little impact on the MA. Under the DBA attack, Multi-Krum almost has no effect. Flame and our method also deteriorate because 40\% of the participants in each round are malicious, which significantly impacts the method of selecting multiple clients for aggregation. Although Krum and Foolsgold successfully defend against the DBA attack, they also negatively affect the overall performance of the global model. Our defense can reduce the BA to less than $10\%$ without affecting the MA. 

PGD attack is much stealthier than the above two which  Krum, RFA, and Multi-Krum fail to detect. Moreover, once they select the wrong gradient for aggregation, the global model is replaced, leading to a worse BA than FedAvg. Ours and Flame achieve the best effect of $0.56\%$ and $0.00\%$ on the two datasets, respectively. Edge-case PGD is the hardest of all since it is both stealthy and effective since it attacks with edge-case samples and projects the gradient back to the $L_2$ norm ball. All the previous defenses have limited effect on this attack. Flame reduces the BA to $5.12\%$. However, Flame suffers from a $6\%$ drop on MA, which is a result of adding noise. In comparison, ours achieves a much lower BA of $3.06\%$ with almost no impact on MA. 

On EMNIST, we also achieve better defense compared to up-to-date methods. Remarkably, we achieve $0.0\%$ BA under both PGD and Edge-case PGD attacks meaning that we defend entirely against these two adversaries.
We also achieve similar performance without sacrificing the benign performance against model replacement and DBA. To provide an overall comparison with previous methods, we follow \cite{ye2022ood} and design an improved version of the ranking score for each defense with respect to baseline FedAvg. We use the percentage of improvement over the baseline to score and compute it on MA/BA respectively, formally as $\frac{K_{\text{MA/BA}}-B_{\text{MA/BA}}}{B_{\text{MA/BA}}}$ where $K$ denotes some methods and $B$ denotes baseline. Adding up $(score_{MA}-score_{BA})$ across all attacks produces the ranking score for each defense. As shown in the last column of Table \ref{compare_table}, we show that our method obtains the highest ranking score with almost 400\% better than the baseline and outperforms the second-best Flame by around $0.5$. We detail the ranking score in Appendix \ref{ranking_score}.
\begin{figure}[htbp]
\centering
\includegraphics[width=3.25in]{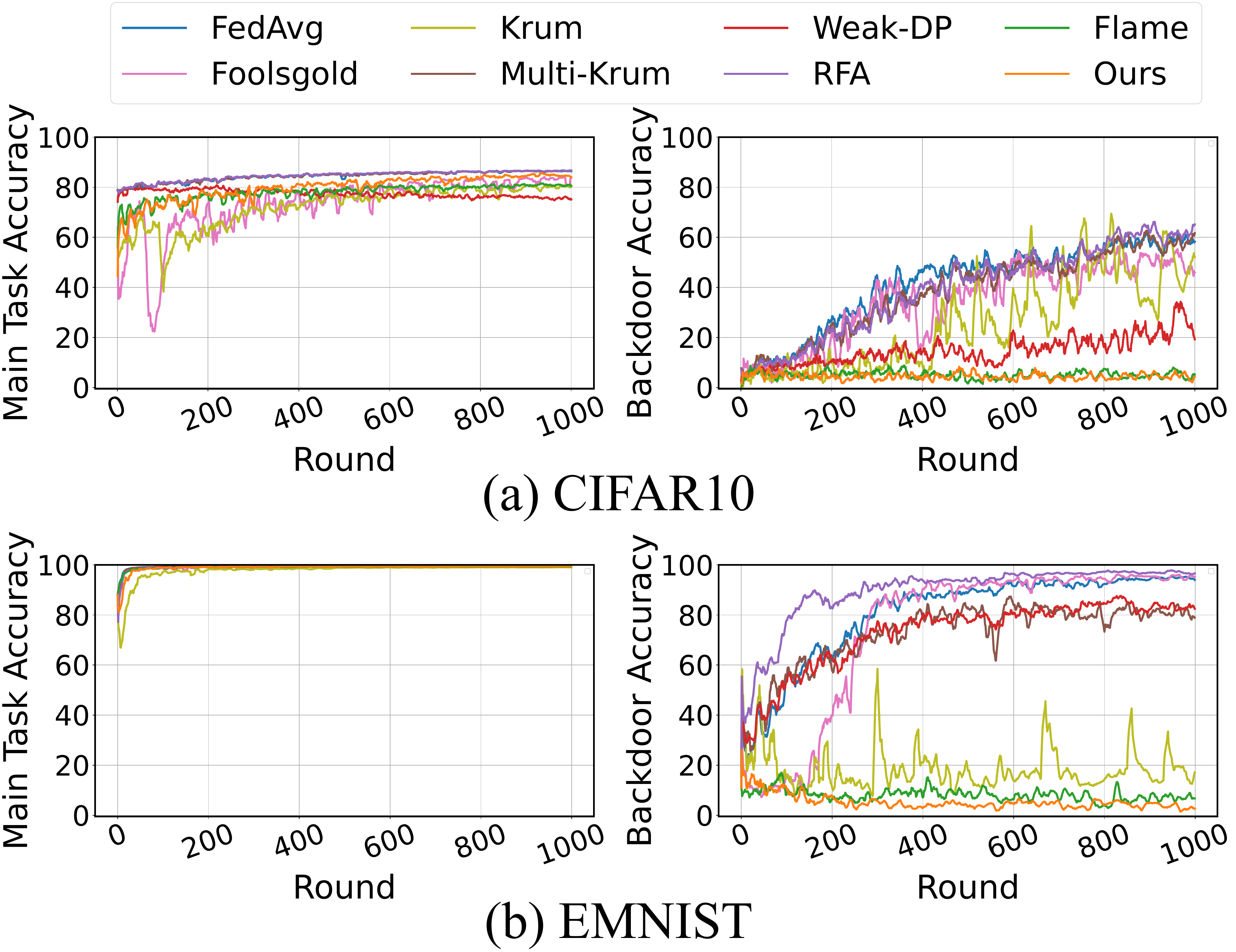}
\caption{MA(\%) and BA(\%) of various defenses under Edge-case PGD attack.}
\label{compare}
\end{figure}

We show the MA and BA along the training process of the seven defenses under the Edge-case PGD attack in Figure \ref{compare}. Only our method and Flame successfully resist the attack during the entire training process, and Flame also dampens the MA.
The disadvantage of our approach is that the convergence speed is slightly slowed down compared to FedAvg, but it is still much better than the compared defenses, \eg, Krum and Foolsgold.

\subsection{Impact of Training Environment}
In this section, we demonstrate that our defense is invariant to the different training environments, \ie degrees of non-IID, and attack frequency.

\begin{figure}[htbp]
\centering
\label{noiid_cifar}
\includegraphics[width=3.25in]{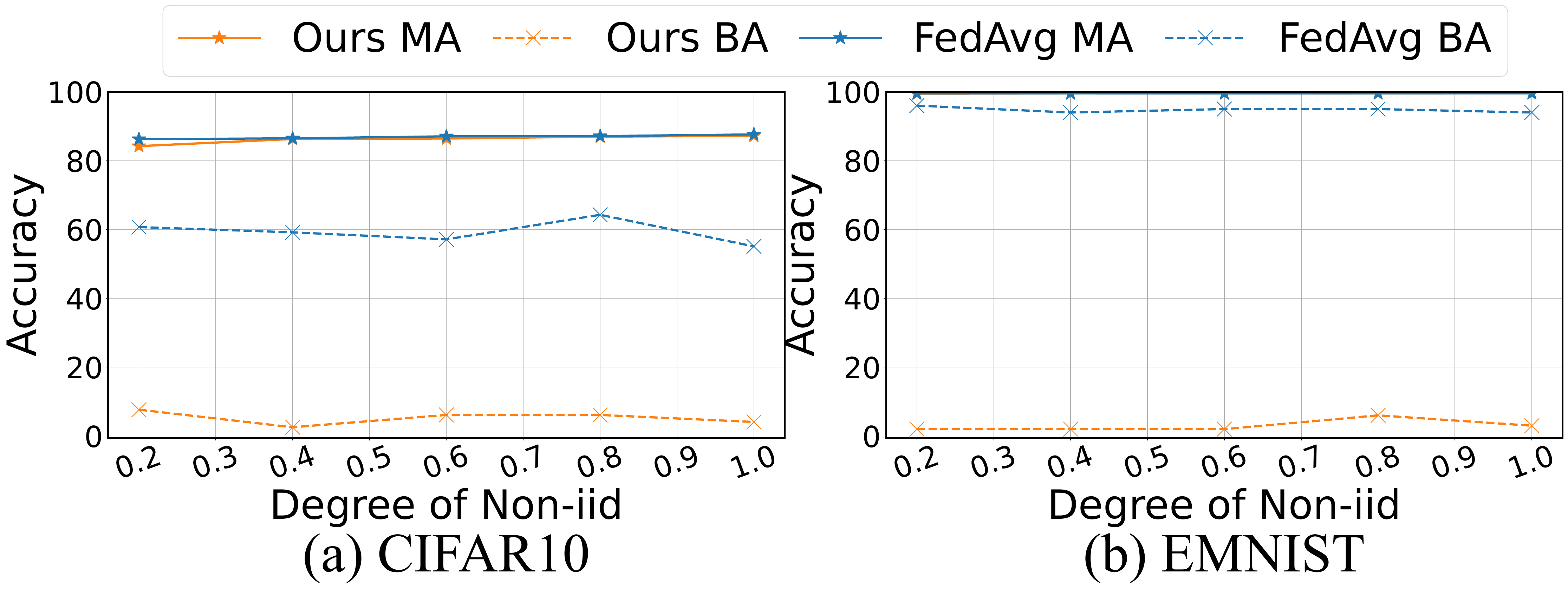}
\caption{Accuracy(\%) of our defense under Edge-case PGD attack on different degrees of non-IID, which is indicated by the $a$ parameter of the Dirichlet distribution.}
\label{noiid}
\end{figure}

\textbf{Impact of Different Degrees of Non-IID.}
As mentioned abundantly in Section \ref{weighting_section}, the different data distributions contribute to one of the hardest problems in distance-based defense. Thus, to illustrate the effectiveness of our method, we demonstrate that our defense is invariant to the different degrees of non-IID. The experiment is conducted with the Edge-case PGD attack. Figure \ref{noiid} shows that the BA remains low in all cases. Our method only slightly impacts the MA with $a=0.2$.


\begin{figure}[htbp]
\centering
\includegraphics[width=3.25in]{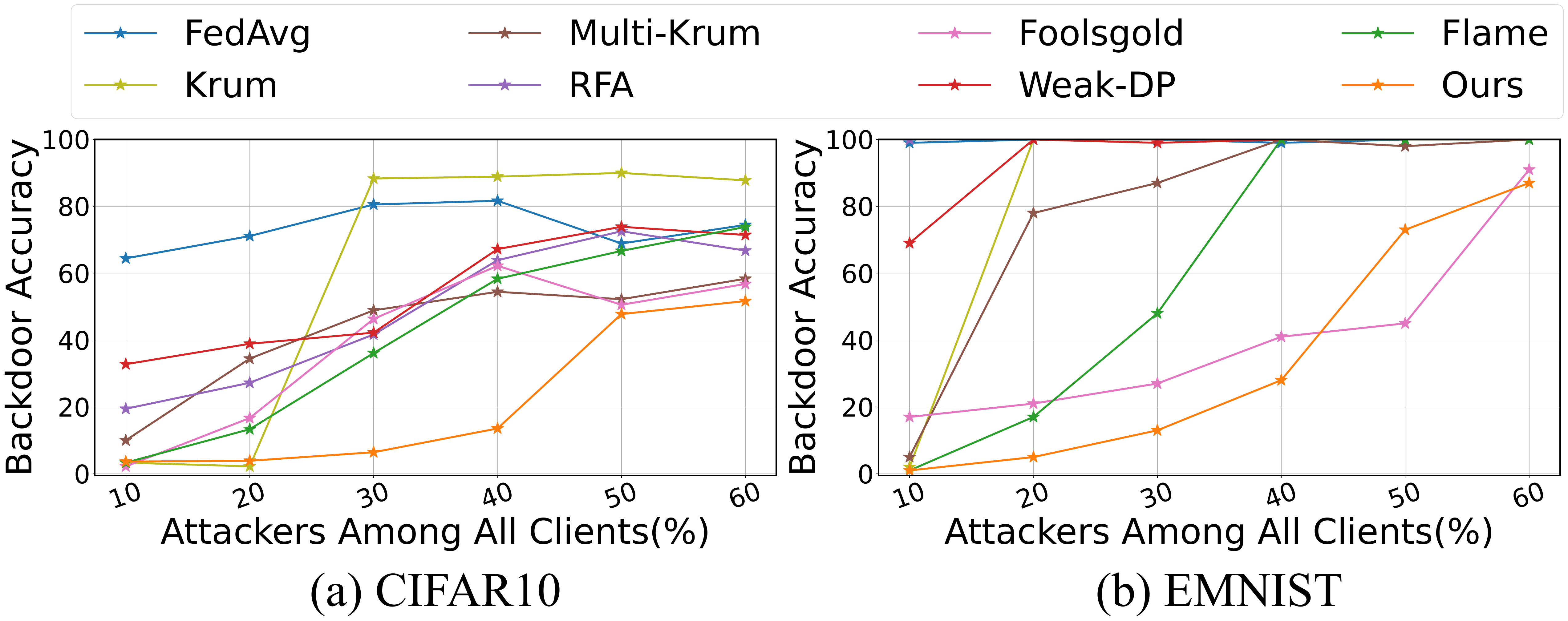}
\caption{BA(\%) of various defenses under the backdoor attack versus percentage of attackers among all clients.}
\label{compare_pool}
\end{figure}

\textbf{Impact of Attacker Percentage.}
We show the BA performance with different ratios of attackers under the backdoor attack in Figure \ref{compare_pool}, where we compare ours with other methods. As a result of our rigorous experiment settings, it usually selects more than $K/2$ adversarial models in each round, so all defenses, especially the clustering-based methods, fail to defend when 50\% of clients are compromised. For EMNIST, BA rises steeply as the simplicity of the task, achieving high accuracy with few parameters injected. Nonetheless, our method performs significantly superior to other methods, and exhibits invariance to the number of attackers encountered.
\begin{figure}[htbp]
\centering
\label{interval_cifar}
\includegraphics[width=3.25in]{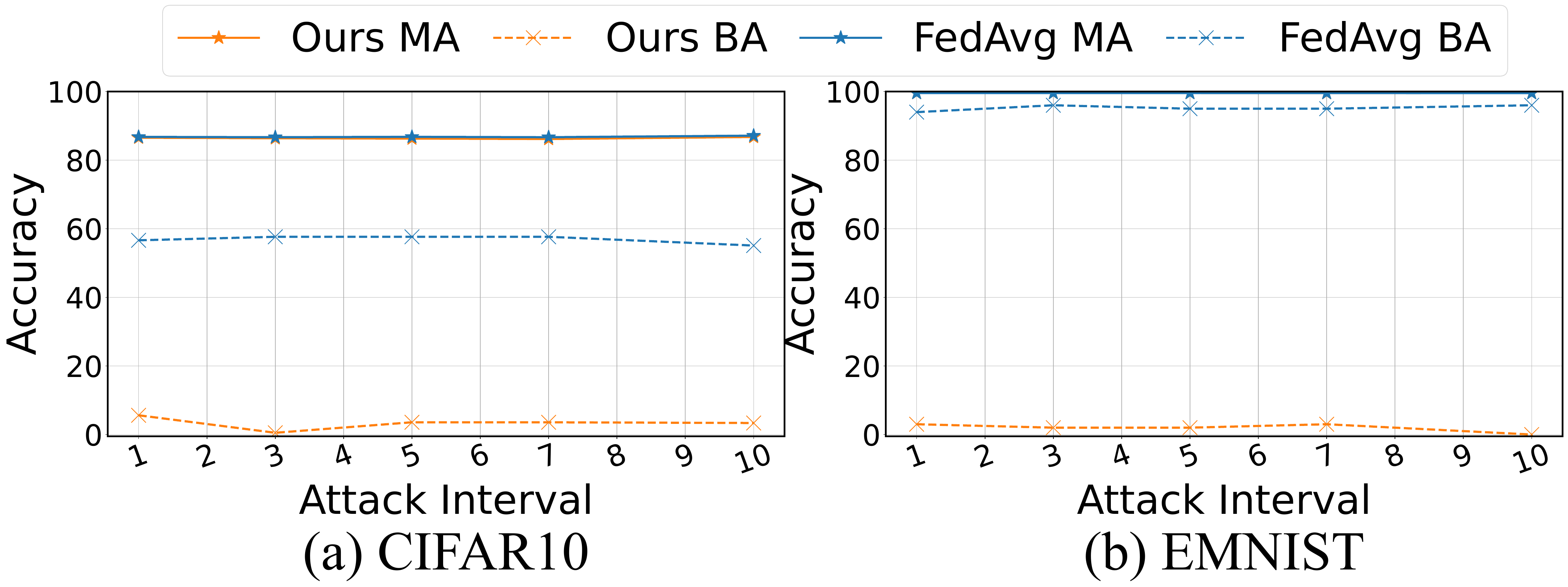}
\caption{Accuracy(\%) of our defense under Edge-case PGD attack versus attack interval.}
\label{interval}
\end{figure}

\textbf{Impact of Attack Frequency.}
The frequency of the attack affects the strength and persistence of the attack. We demonstrate that our defense is invariant to the frequency of attack in Figure \ref{interval} where we show that our defense achieves consistently low BA under all attack intervals. Since the other defenses cannot resist the edge-case PGD with the minimum frequency(attack interval of 10), we do not plot their results.

\subsection{Generalization to Different Datasets}
We experiment on the LOAN and Sentiment140, as shown in Table \ref{general}. Our method achieves high MA and low BA on both datasets, proving the adaptivity to different models and tasks.

\begin{table}[htbp]
\caption{Effectiveness of our method against backdoor attack on other types of datasets.}
\label{general}
\centering
\resizebox{0.42\textwidth}{!}{
    \begin{tabular}{@{}ccccccc@{}}
    \toprule
    \multirow{2}{*}{Defense} & \multicolumn{2}{c}{CINIC10}    & \multicolumn{2}{c}{LOAN}    & \multicolumn{2}{c}{Sentiment140} \\ \cmidrule(l){2-7} 
                             & MA $\uparrow$            & BA $\downarrow$       & MA $\uparrow$            & BA $\downarrow$        & MA $\uparrow$             & BA $\downarrow$        \\ \midrule
    FedAvg                   & \textbf{80.02} & 36.22         & \textbf{89.05} & 61.36      & \textbf{82.59}  & 89.17          \\
    Ours                     & 76.24          & \textbf{4.59} & 88.52          & \textbf{0} & 81.67           & \textbf{5.83}  \\ \bottomrule
    \end{tabular}
}
\end{table}

\subsection{Ablation Study}
\label{ablation study}
The above experiments present the remarkable defense ability of our approach. In this section, we conduct an ablation study to understand the role of each component.

\begin{figure}[htbp]
\centering
\includegraphics[width=2.7in]{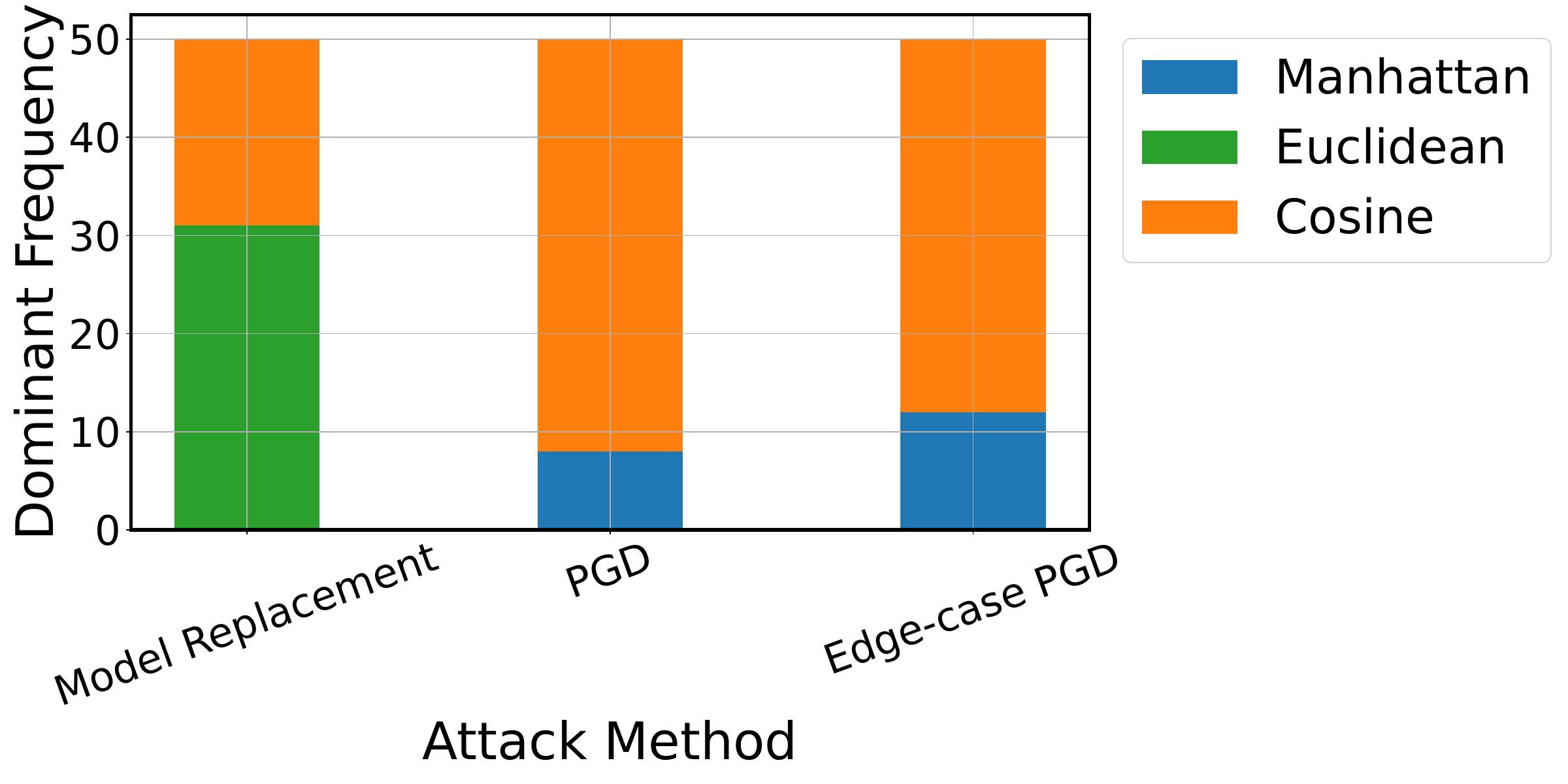}
\caption{Frequency of dominance of each metric under different attacks on CIFAR10.}
\label{weight}
\end{figure}

\textbf{Multi-metrics Identifies the Stealthiest Attacks.} We first conduct an ablation study on the multi-metrics, as shown in Table \ref{metric_performance}. We observe the following: 1. All three metrics perform poorly individually. 2. Manhattan distance improves the defense ability when combined with the other metrics. 3. Cosine similarity combined with Manhattan distance provides the best defense against the Edge-case PGD. Given the above observations, we conclude that multi-metrics adaptively defends against universal backdoors. The introduced Manhattan metric contributes to the defense against the stealthiest attack (\eg, Edge-case PGD).
\begin{table}[]
\centering
\caption{Effectiveness of the metrics in our approach to defending against the various attacks on CIFAR10.}
\label{metric_performance}
\resizebox{0.46\textwidth}{!}{
\begin{tabular}{@{}cccc@{}}
\toprule
\multirow{2}{*}{Defenses} & \begin{tabular}[c]{@{}c@{}}Model\\ Replacement\end{tabular} & PGD         & \begin{tabular}[c]{@{}c@{}}Edge-case\\ PGD\end{tabular} \\ \cmidrule(l){2-4} 
                          & MA/BA                                                       & MA/BA       & MA/BA                                                   \\ \midrule
Man                       & 83.86/\textbf{0.56}                                                  & 83.74/25.56 & 85.3/64.80                                              \\
Eul                       & 86.24/\textbf{0.56}                                                  & 85.52/17.78 & \textbf{87.12}/54.08                                             \\
Cos                       & 84.22/2.22                                                  & 83.84/30.00    & 85.38/66.84                                             \\
Man+Eul                   & 84.09/1.11                                                  & 84.17/28.63 & 84.3/67.35                                              \\
Man+Cosine                & 85.74/1.67                                                  & 85.16/23.68 & 85.86/6.63                                              \\
Cosine+Eul                & 86.31/\textbf{0.56}                                                  & 85.44/16.11 & 85.14/63.78                                             \\
Man+Cosine+Eul            & \textbf{86.34}/\textbf{0.56}                                                  & \textbf{86.44}/\textbf{0.56}  & 86.86/\textbf{3.06}                                              \\ \bottomrule
\end{tabular}
}
\end{table}
To further demonstrate the claim that a single metric cannot identify malicious gradients of diverse characteristics, we show the dominance of each metric under different attacks. 
We take the metric as the dominant one with the most considerable contribution to this round's scoring and plot their frequency in Figure \ref{weight}. 

Firstly, we show that Manhattan distance contributes to the defense of PGD and Edge-case PGD, complementing the Cosine similarity justifying the above observations 2 and 3. We also notice that Euclidean is never dominant under PGD and Edge-case PGD, consistent with the analysis that these two attacks bypass $L_2$ norm through projection back to $L_2$ norm ball. 
We also notice that Manhattan distance plays no role under the model replacement attack, consistent with the above experiment that Manhattan does not affect the final BA. It is because model replacement attack scales the gradient of the malicious client, making it susceptible to Cosine and Euclidean distances.
Finally, we conclude that better defense can be achieved if multiple instead of single metrics are utilized since different metrics dominantly detect different malicious gradients.

\textbf{Dynamic Weighting through Whitening Process Adaptively Adjust the Scoring.} 
We illustrate the effectiveness of our whitening process discussed in Section \ref{weighting_section} in Table \ref{weight_compare} by comparing it with the standard max normalization. BA increases by a large margin if only max normalization is applied, which illustrates that the whitening process plays an essential role in identifying the malicious under non-IID distributions. Moreover, we notice the MA increases as we apply the whitening process, which shows that it can flexibly identify the benign but different gradients (due to non-IID) and the true malicious ones.
\begin{table}[htbp]
\centering
\caption{Effectiveness of different weighting methods on defense against various attacks on CIFAR10.}
\label{weight_compare}
\resizebox{0.49\textwidth}{!}{
\begin{tabular}{@{}ccccccc@{}}
\toprule
                                                             & \multicolumn{2}{c}{Model Replacement} & \multicolumn{2}{c}{PGD} & \multicolumn{2}{c}{Edge-case PGD} \\ \cmidrule(l){2-7} 
\multirow{-2}{*}{Weighting}                              & MA $\uparrow$                 & BA $\downarrow$               & MA$\uparrow$         & BA $\downarrow$         & MA $\uparrow$              & BA $\downarrow$              \\ \midrule
Max Norm                                                        & 83.86              & \textbf{0.56}             & 83.74      & 25.56      & 84.08           & 62.24           \\
Whitening                                                 & \textbf{86.34}              & \textbf{0.56}             & \textbf{86.44}      & \textbf{0.56}       & \textbf{86.86}           & \textbf{3.06}            \\ \bottomrule
\end{tabular}
}
\end{table}

\section{Conclusion}
Existing defense methods fail to handle backdoor attacks with malicious updates similar to benign ones, especially when benign clients possess data of non-independent and iden
tical distribution. In this paper, we propose a novel adaptive multi-metrics defense method by leveraging multiple metrics with dynamic weighting to defend against backdoor attacks in FL, withstanding a wide range of stealthy and elaborate attacks. We conduct exhaustive experiments to evaluate our approach proving its effectiveness over a wide range of attack settings.

\section*{Acknowledgements} 
This work is supported by the Key-Area Research and Development Program of Guangdong Province under Grant 2019B010137001.

{\small
\bibliographystyle{ieee_fullname}
\bibliography{egpaper_for_review}
}

\appendix
\newpage

\section{Additional expertiment}
\subsection{Impact of Different Fraction of Clients Selected for Aggregation}
\label{p_ablation}
\begin{figure}[htbp]
\centering
\includegraphics[width=3.1in]{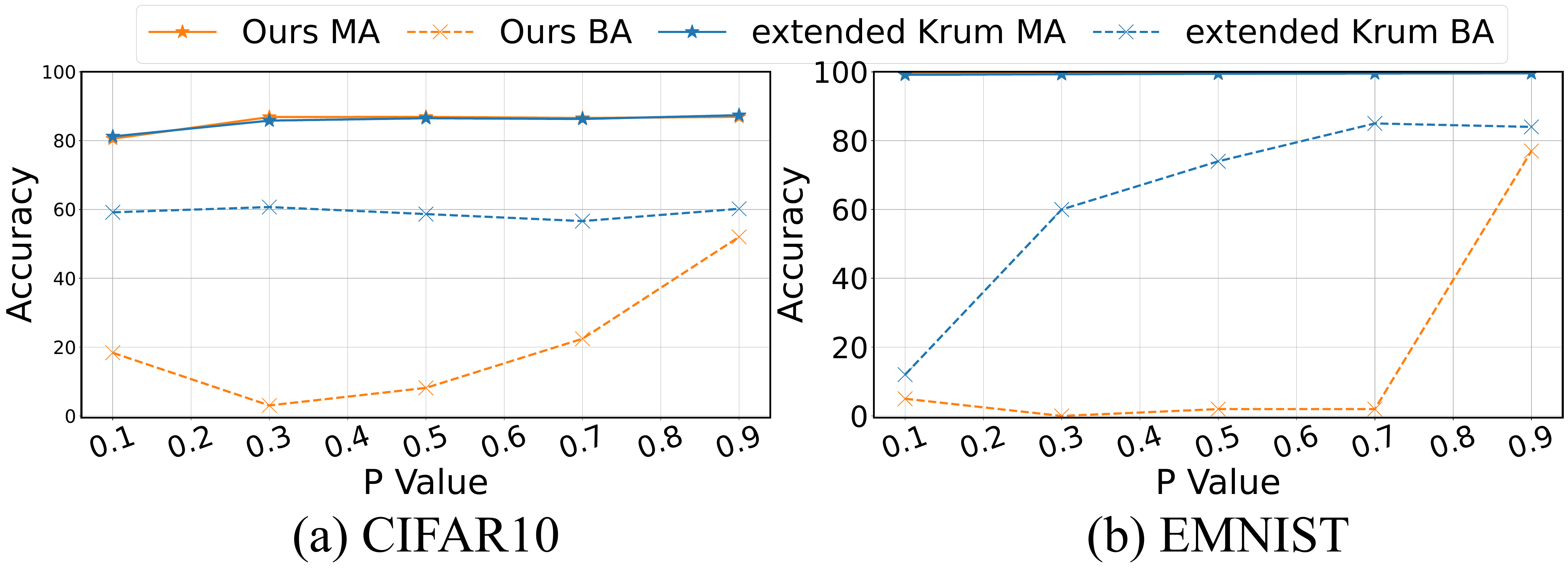}
\caption{Accuracy(\%) of our defense and extended-Krum under Edge-case PGD attack versus the value of $p$, where $p$ denotes the fraction of clients selected for aggregation by our method.}
\label{p_value}
\end{figure}

In each round of federated training, our method distinguishes the benign gradients from the malicious ones through  multiple metrics and dynamic scoring. Then benign gradients are used to perform aggregation while the malicious ones are discarded without impacting the global model, as described in Section \ref{p_section}. In practice, we set a fixed ratio $p(p \in [0,1])$ to denote the fraction of the selected gradients. At each round, $p$ percentage of gradients are deemed benign and participate in the FedAvg aggregation. Intuitively, the performance of the model and the convergence speed of training are positively related to $p$. In contrast, the relationship between the accuracy of backdoor tasks and $p$ is much more complicated. On the one hand, increasing the value of $p$ would increase the probability of selecting the backdoor gradient for training, which is not beneficial for defending against backdoor attacks. On the other hand, increasing the value of $p$ will mitigate the impact of selecting the backdoor gradient, which is beneficial for the defense. However, because of an absence of knowledge regarding the attacker (\eg the number of attackers), optimal $p$ cannot simply be determined. In this case, what is most essential is that the defense performance is invariant (or often invariant) to the choice of $p$, which we empirically prove below. By conducting experiments on CIAFR10 and EMNIST with the Edge-case PGD attack, we show that our outstanding defense performance does not heavily rely on the tuning of $p$. Results in Figure \ref{p_value} show that the optimal $p$ is at $0.3$ but $p$ between $0.1$ and $0.7$ provides consistently low BA, where our approach has solid defensive effectiveness. Krum and Multi-Krum also select some clients to aggregate, which we make a comparison in terms of defense performance with our proposed method under different $p$. We demonstrate that our success against this attack does not rely on an optimal $p$ and consistently gives better results against the Krum and Multi-Krum with different $p$ values.

\subsection{Ablation on Different Definitions for Outlier Detection}
We also conduct an ablation study on the definition, as introduced by Equation \ref{redefinition}. We perform experiments and compare the results of the simple approach (\ie use the deviation from the mean to detect the outlier as a malicious gradient). As the results in Table \ref{definition_compare}, we observe that the simple approach is already effective under different attacks. Specifically, with a simple approach, our defense can achieve $7.14\%$ BA under Edge-case PGD which already outperforms the previous state-of-the-art methods by almost $40\%$, demonstrating the effectiveness of our multi-metrics adaptive defense method. By incorporating the new definition, we achieve the lowest $3.06\%$. Results under other attacks are also consistent. We note that our new definition increases the overall performance (\ie MA) under the three attacks consistently. This demonstrates that our defense can aggregate the truly benign gradients since this new definition helps further identify the malicious gradients.

\begin{table}[htbp]
\caption{Impact of different definitions for outlier detection against various attacks on CIFAR10.}
\label{definition_compare}
\resizebox{0.46\textwidth}{!}{
\begin{tabular}{@{}ccccccc@{}}
\toprule
\multirow{2}{*}{Defense} & \multicolumn{2}{c}{Model Replacement} & \multicolumn{2}{c}{PGD} & \multicolumn{2}{c}{Edge-case PGD} \\ \cmidrule(l){2-7} 
                         & MA $\uparrow$                 & BA $\downarrow$               & MA $\uparrow$          & BA $\downarrow$        & MA $\uparrow$               & BA $\downarrow$             \\ \midrule
Mean       & 85.91              & \textbf{0.56}             & 85.83       & 1.67      & 85.82            & 7.14           \\
Ours       & \textbf{86.34}              & \textbf{0.56}             & \textbf{86.44}       & \textbf{0.56}      & \textbf{86.86}            & \textbf{3.06}           \\ \bottomrule
\end{tabular}
}
\end{table}


\begin{table}[hbtp]
\centering
\caption{Computational measured in seconds. We report the increment over FedAvg.}
\label{computational_cost}
\resizebox{0.49\textwidth}{!}{
\begin{tabular}{@{}ccccc@{}}
\toprule
Defense                & \textbf{FedAvg} & Krum         & RFA                & Foolsgold     \\ \midrule
Computational Cost (s) & \textbf{24.62}  & 25.98(+1.36) & 31.08(+6.46)       & 78.67(+54.05) \\ \midrule
Defense                & Single metric    & Dual metrics  & Tri metrics (Ours) & Four metrics  \\ \midrule
Computational Cost (s) & 25.55(0.93)     & 26.14(1.52)  & 30.19(+5.57)       & 33.74(+9.12)  \\ \bottomrule
\end{tabular}
}
\end{table}

\begin{table*}[t]
\centering
\caption{The training settings for our experiment.}
\label{parameters}
\begin{tabular}{@{}ccccccc@{}}
\toprule
\multirow{3}{*}{Hyperparameter}    & \multicolumn{6}{c}{backdoor type}                                         \\ \cmidrule(l){2-7} 
                                   & \multicolumn{3}{c}{semantic}                & \multicolumn{3}{c}{trigger} \\
                                   & CIFAR10            & EMNIST & Sentiment140    & CIFAR10  & EMNIST  & LOAN    \\ \midrule
backdoor data                      & Southwest Airlines & Ardis & Greek Director & N/A      & N/A    & N/A     \\
\#clients                          & 200                & 200   & 2000           & 100      & 100    & 100     \\
\#clients selected in each   round & 10                 & 10    & 10             & 10       & 10     & 10      \\
\#attackers in each round          & 1                  & 1     & 1              & 4        & 4      & 4       \\
\#attackers local iteration        & 5                  & 5     & 2              & 6        & 10     & 5       \\
\#benign local iteration           & 2                  & 2     & 2              & 2        & 1      & 1       \\
\#global iteration                 & 1500               & 1500  & 200            & 300      & 70     & 70      \\
batch size                         & 64                 & 64    & 20             & 64       & 64     & 64      \\
attack interval                    & 10                 & 10    & 10             & 1        & 1      & 1       \\
no.iid parameter                   & 0.5                & 0.5   & 1              & 0.5      & 0.5    & 0.9     \\
benign learning rate               & 0.02               & 0.02  & 0.05           & 0.1      & 0.1    & 0.001   \\
attackers learning rate            & 0.02               & 0.02  & 0.05           & 0.05     & 0.05   & 0.0005  \\ \bottomrule
\end{tabular}
\end{table*}

\subsection{Computational Cost Analysis}
We acknowledge that there is a slight increase in computation overhead compared to FedAvg, we emphasize that the additional cost is marginal and significantly lower than the previous SOTA, Foolsgold, as shown in Table 
\ref{computational_cost}.

\section{Training Hyperparameters}
\label{hyperparameters}
The pixel-pattern backdoor data used in the DBA attack is the same as that in Xie et al. \cite{xie2019dba}. Following \cite{wang2020attack}, we use the data from Southwest Airlines as the dataset for the semantic backdoor attack. Particularly, edge-case indicates that the backdoor data exists only in the attacker's dataset. For a normal attack (non-edge-case), we distribute $10\%$ of the correctly labelled backdoor data to benign clients. 

The hyperparameters in FL system can be seen in Table \ref{parameters}, which used in defenses is as follows:
\begin{itemize}
    \item Multi-krum:In our experiment, we select the hyperparameter $m = n - f$ (where $n$ stands for the number of participating clients and $f$ stands for the number of tolerable attackers);
    \item RFA:We set $v = 10^{-5}$ (smoothing factor), $\varepsilon = 10^{-1}$(fault tolerance threshold), $T = 500$ (maximum number of iterations);
    \item Weak-DP:In our experiment, we use $\sigma = 0.0025$ and set the norm difference threshold at $2$.
    \item Flame: In our experiment, we follow the original paper and use small noise  $\sigma = 0.001$.
\end{itemize}

\section{Proof of Proposition \ref{prop}}
\label{proof_prop}
From the Lemma \ref{lem}, we can compute the value of $M_d$ and $U_d$ approximately below, where $M_d=Dmax_d^1-Dmin_d^1$ reflects the discriminating ability of Manhattan distance and $U_d=Dmax_d^2-Dmin_d^2$ reflects the discriminating ability of Euclidean distance.

\begin{equation}
\label{proof_M}
\lim _{d \rightarrow \infty} E\left[ \frac{M_d}{d^{\frac{1}{2}}}\right] = C_1.
\end{equation}
where C1 is a constant.

\begin{equation}
\label{proof_U}
\lim _{d \rightarrow \infty} E\left[ U_d\right] = C_2.
\end{equation}
where C2 is a constant.
Thus, we can divide them to compare the $M_d$ and $U_d$ as follows:
\begin{equation}
\lim _{d \rightarrow \infty} E\left[\frac{M_d}{U_d \cdot d^{\frac{1}{2}}}\right] = \frac{C1}{C2} = C'
\end{equation}
where C' is a constant.

\section{Ranking Score}
\label{ranking_score}
To facilitate a comprehensive comparison that involves multiple attack methods and two metrics (\ie MA and BA), we base on the ranking score in \cite{ye2022ood} and design a new scoring rank that considers both the two metrics and the relative improvement over baseline. The original ranking score used in \cite{ye2022ood} aims to compare the performance over multiple domain generalization datasets while we want to compare the defense over a set of attacks. They first set a baseline method and for every dataset-algorithm pair, depending on whether the attained accuracy is lower or higher than the baseline accuracy on the same dataset,  -1 or +1 is assigned and they add up the scores across all datasets to produce the ranking score for each algorithm. \textbf{Why don't we use the original ranking score?} The problem with the original score is, despite using +1 and -1 to indicate higher or less, the relative improvement or decrease is not explicitly accounted for in the final score, which can be problematic in our setting as defense performance varies a lot across different attacks (from $\sim 10\%$ to  $\sim 90\%$) and scales of the two metrics are also significantly different (from $\sim 30\%$ to $\sim 80\%$). For example, if one method obtains a small improvement (\eg 1\%) on MA, but a much worse BA (decrease by 30\%), the original score method gives a +1 and -1 which makes this method as good as the baseline. However, sacrificing a little MA is still a good price to pay for a large BA improvement. Thus, we propose our new ranking score as followed. We first set a baseline method, \ie FedAvg. Then for each defense method, each attack and each metric, we calculate the relative improvement with respect to the baseline:
$$\text{score}_{K} = \frac{K-B}{B}$$
where $K$ denotes the MA or BA of some methods and $B$ denotes the MA and BA of the baseline. Adding up the score for MA and BA 
$$\text{score}_{\text{MA}}-\text{score}_{\text{BA}}$$
across all attacks produces the ranking score for each defense. Note here we use subtraction instead of addition since a smaller BA means better  defense. Another benefit of this metric is that we can show an average of relative improvement over the baseline, which gives us a sense of how good each method is. We compare our defense with previous SOTA on this metric and provide a detailed analysis in Section \ref{compare_defenses}.

\end{document}